# Spectroscopic Characterization of Landau Level Splitting and the Intermediate $v = 0$ Phase in Bilayer Graphene

Long-Jing Yin[1,2], Li-Juan Shi[2], Li-Zhen Yang[2], Ling-Hui Tong[2], and Lin He[1]*

[1] *Center for Advanced Quantum Studies, Department of Physics, Beijing Normal University, Beijing, 100875, China*

[2] *Key Laboratory for Micro/Nano Optoelectronic Devices of Ministry of Education & Hunan Provincial Key Laboratory of Low-Dimensional Structural Physics and Devices, School of Physics and Electronics, Hunan University, Changsha 410082, China*

*Corresponding author: helin@bnu.edu.cn

**Despite various novel broken symmetry states have been revealed in bilayer graphene (BLG) experimentally, the atomic-scale spectroscopic investigation has been greatly limited. Here, we study high-resolution spectroscopic characteristics of high-quality BLG and observe rich broken-symmetry-induced Landau level (LL) splittings, including valley, spin and orbit, by using ultralow-temperature and high-magnetic-field scanning tunneling microscopy and spectroscopy (STM and STS). Our experiment demonstrates that both the spin and orbital splittings of the lowest $n = (0,1)$ LL depend sensitively on its filling and exhibit an obvious enhancement at partial-filling states. More unexpectedly, the splitting of a fully-filled and valley-polarized LL is also enhanced by partial filling of the LL with the opposite valley. These results reveal significant many-body effects in this system. At half filling of the $n = (0,1)$ LL (filling factor $v = 0$), a single-particle intermediate $v = 0$ phase, which is the transition state between canted antiferromagnetic and layer-polarized states in the BLG, is measured and directly visualized at the atomic scale. Our atomic-scale STS measurement gives direct evidence that this intermediate $v = 0$ state is the predicted orbital-polarized phase.**

Due to the multiple electronic degrees of freedom, bilayer graphene (BLG) is emerging as a fascinating experimental platform for exploring rich novel phenomena in the quantum Hall regime. The coexistence of spin (↑↓), valley ($K/K'$), and orbital ($n = 0,1$) quantum numbers generates an eightfold degeneracy of the lowest Landau level (LL) in the BLG [1-4]. The SU(8) symmetry of the lowest LL can be broken by single-particle effects such as application of an interlayer electric field and/or by Coulomb interactions controlled by magnetic fields, giving rise to the emergence of various broken symmetry states and a complex phase diagram [5-21]. Experimentally, great successes have been achieved in revealing a sequence of novel broken symmetry states in the lowest LL by several techniques, mainly including transport [8-17], capacitive [18,19] and single-electron transistor [20,21] measurements. However, the spatially resolved spectroscopic detection of those novel broken symmetry states has been greatly limited so far.

In this letter, we present scanning tunneling microscopy and spectroscopy (STM and STS) measurements of high-quality BLG on graphite substrate under ultralow-temperature (~500 mK) and high magnetic field (up to 15 T). There are two unique advantages of the STM/STS measurements: i) the atomic-scale spatial resolution [22-24], which allows us to visualize the quantum phases in real space and, therefore, directly identify them [25,26]; ii) the ability to probe the electronic states both near and far from the Fermi level, which enables us to investigate the splitting of the broken symmetry states at different filling, ranging from full filling to complete empty, in the BLG. Our result demonstrates that both the spin and orbital splittings of the lowest $n = (0,1)$ LL depend sensitively on the filling state and have an enhancement at partial-filling, revealing remarkable many-body effects in the lowest LL of the BLG. At half filling of the $n = (0,1)$ LL (filling factor $v = 0$), an intermediate transition $v = 0$ phase, which has attracted much attention very recently [4,5,14,16,18,27], is detected and directly visualized at the atomic scale in our experiment. Our atomic-scale measurement demonstrates that this intermediate $v = 0$ state is the predicted orbital-polarized phase.

Our BLG samples are electronically decoupled bilayer graphene on graphite substrate (see Supplemental Material [28] for details of the sample preparation and measurements). The decoupled *AB* (Bernal) stacked BLG flakes can be well identified by the STM measurement [Fig. 1(a)] combined with magnetic-field-varied STS spectra [Figs. 1(b) and (c)]. Figure 1(a) shows an atomic-resolution STM topographic image of an *AB*-stacked BLG. For the *AB*-stacked BLG, the *B* sublattice of the top layer ($B_T$) is right above the *A* sublattice of the bottom layer ($A_B$), and the top *A* sublattice ($A_T$) is located at the center of the hexangular holes in the underlying layer. Due to the direct coupling between the $B_T$ and the $A_B$, the density of states (DOS) of the $B_T$ is suppressed and the $A_T$ is dominantly visualized, leading to the triangular structure in the STM topographic image [see Fig. 1(a)] [29,30]. Figure 1(b) shows a representative zero-field STS spectrum, *i.e.*, *dI/dV* spectrum, of the sample. There are four peaks in the tunneling spectrum: two peaks are located near zero bias (P1 and P2) and the other two are at ~ ±0.3 V (P3 and P4). The two peaks at ~ ±0.3 V are generated by the edges of the high-energy bands in the BLG [see left inset of Fig. 1(b)], which are rarely observed in previous experiment, indicating the high-quality of the studied sample. The two low-energy peaks are generated by the DOS peaks at the rather flat valence- and conduction-band edges of a gapped BLG. The band gap is of ~28 meV [see right inset of Fig. 1(b)], which is generated by a substrate-induced interlayer potential [31-33]. The two low-energy states are localized in different layers, resulting in the large asymmetry of their intensities [34,35]. The full width at half maximum of P1 is only ~3.4 meV, meaning that the conduction-band edge is almost flat. This ~3.4 meV band width is even smaller than that of the flatbands measured in magic-angle twisted bilayer graphene (~18 meV) [36] and in ABC-stacked trilayer graphene (~6 meV) [37], suggesting that exotic strongly correlated phenomena, such as correlated insulator and superconductivity, could also be realized in the BLG [38,39].

Figure 1(c) shows *dI/dV* spectra of the gapped BLG measured at various magnetic fields. The high-magnetic-field tunneling spectra exhibit a series of well-separated LLs, indicating that our BLG sample is of high-quality and is effectively decoupled from the substrate. We can extract band structure information of the BLG from the measured LL spectra. The semi-classical Onsager quantization condition provides an equation for the $k$-space area $S_n$ of the $n$th LL at energy $E_n$: $S_n = (n+\gamma)2\pi eB/\hbar$ ($\gamma$ is the phase offset and is zero for Dirac fermions) [40,41]. At low energy, the constant-energy contour is almost circular, $S_n$ can be taken as $S_n = \pi k_n^2$, and thus we have $k_n = \sqrt{2enB/\hbar}$. Figure 1(d) shows the measured energy position $E_n$ and its associated $k_n$ for 4-10 T, which traces out an effective dispersion relation of a gapped BLG. The band parameters are determined by fitting the experimental data with the tight-binding model and we obtain $\gamma_0 = 2.65$ eV, $\gamma_1 = 0.31$ eV, $\gamma_3 = 0.1$ eV, $\gamma_4 = 0.08$ eV, and $u = 28$ meV (see Fig. S1 [28] for details), which are consistent well with that reported previously [42].

Besides the LLs, the high-resolution spectra in Fig. 1(c) exhibit several other notable features. First, rich LL splittings, including $n = (0,1)$, $n = 2$ and $n = 3$ orbits, are observed. Among these, the splitting of the $n = (0,1)$ LL is the most remarkable. The valley degeneracy of the $n = (0,1)$ LL is lifted first due to the existence of an interlayer potential, resulting in two layer-polarized quartets: $LL_{(0,1,+)}$ and $LL_{(0,1,-)}$ [as marked in Fig. 1(c). Here +/- denotes two valleys]. The wavefunctions of electrons from the $LL_{(0,1,+)}$ are localized on the $A_T$, whereas the $LL_{(0,1,-)}$ states are predominantly on the $B_B$. This layer polarization results in that the peak of the $LL_{(0,1,+)}$ is much more intensive than that of the $LL_{(0,1,-)}$ in the STS spectra [Fig. 1(c)], as the STM tip predominantly probing the DOS of the top layer [31-33,35]. The 4-fold degeneracy of the $LL_{(0,1,+)}$ and $LL_{(0,1,-)}$ is further lifted: the $LL_{(0,1,+)}$ splits into two peaks and the $LL_{(0,1,-)}$ splits into four peaks [the $LL_{(0,1,-)}$ is close to the Fermi level, see Fig. 1(e)]. According to the Hund's rule in the BLG [2], the two-peak splitting of the $LL_{(0,1,+)}$ and the largest splitting of the $LL_{(0,1,-)}$ are attributed to broken spin degeneracy, and the further splitting of the $LL_{(0,1,-)}$ is attributed to broken orbital degeneracy. Combined with previous results in transport measurements [16,20], the sequence of symmetry breaking observed in our BLG is schematically shown in the inset of Fig. 2(a). In high magnetic fields, both the $n = 2$ and $n = 3$ LLs are split into two peaks, indicating either the spin or the valley degeneracy is lift [16]. At present, we do not know the exact nature of the observed symmetry-breaking states for the $n = 2$ and $n = 3$ LLs (see Fig. S2 [28] for a detailed discussion).

Figure 2 summaries the splitting energies for the broken symmetry states of the $n =$ (0,1), $n = 2$ and $n = 3$ LLs as a function of magnetic fields. The valley splitting of the $n =$ (0,1) LL is ~ 30 meV, which is mainly controlled by the interlayer potential ($u$ ~ 28 meV from our fitting), leading to its weak dependence on the magnetic fields [Fig. 2(a)] [16,20]. The many-body Coulomb interaction is also expected to contribute to the valley splitting in the $n =$ (0,1) LL, and that may be the reason for the slight increase at $B \geq 11$ T. In the magnetic fields ranging from about 9 T to 13 T, the spin splittings of the fully unoccupied $LL_{(0,1,+)}$ and the fully filled $LL_{(0,1,-)}$ are almost linearly scaled with the magnetic fields [Fig. 2(b)] and we obtain the $g$ factor, derived from the slope of the linear fits, ~ 12 and ~ 21 for the $LL_{(0,1,+)}$ and the $LL_{(0,1,-)}$, respectively. The large spin splitting with $g$ factor, which has already been reported in the BLG in other measurements [16,20,21], is attributed to the interaction-enhanced Zeeman splitting [4]. Similarly, the orbital splittings of the fully filled $LL_{(0,1,-)}$ also increase linearly with magnetic fields and the slopes of the orbital splitting for the two spin-split states are measured as 0.56 meV/T and 0.51 meV/T (or effective $g$ factors of $g_o \approx 10$ and 9) respectively. The obtained orbital splitting is consistent with previous experimental results ~ 0.4 meV/T [16,20] and also agrees with the predicted ~ $\gamma_4 B$ dependence in the BLG [18].

For the magnetic fields $B < 9$ T, an obvious enhancement of both the spin splitting and orbital splitting is observed in the partially filled $LL_{(0,1,-)}$. Such a partial-filling-enhanced splitting, which has been observed in monolayer and trilayer graphene [25,37,43], arises from the exchange interaction, indicating strong many-body effects in the lowest LL of the BLG. It is very interesting to note that the spin splitting in the fully unoccupied $LL_{(0,1,+)}$ is also enhanced when the $LL_{(0,1,-)}$ is partially filled, as shown in Fig. 2(b). Such a phenomenon, which has never reported for the LLs, is quite unexpected because that there is no reason to expect an enhanced splitting of the empty bands when the occupation of the other bands is changed. In the magic-angle twisted bilayer graphene, similar feature is also observed in the low-energy flat bands and is demonstrated to be beyond the description of a weak coupling mean-field picture [44].

Therefore, the above experimental feature indicates that electron-electron interactions play a dominant role in the BLG in the presence of high perpendicular magnetic fields. Besides, an unusual suppression of the spin splitting is observed for both the $LL_{(0,1,+)}$ and $LL_{(0,1,-)}$ at $B > 13$ T [see Fig. 2(b)]. This phenomenon may be related to the emergence of an intermediate phase in the BLG, which will be discussed below (see a detailed discussion in Supplemental Material [28]). Figure 2(c) shows the splittings of $n \geq 2$ LLs, which also exhibit a linear $B$-dependence. Similar as previous results obtained in other graphene systems [37,45], the splitting decreases with increasing the LL index: the slope of the splitting is 0.75 meV/T (or effective $g$ factor ~13) for the $n = 2$ LL and it decreases to 0.53 meV/T (or effective $g$ factor ~9) for the $n = 3$ LL.

Another notable feature observed in our experiment is the strong dependence of the relative intensity of the two spin-split peaks in the $LL_{(0,1,+)}$ on the filling states of the BLG. Figures 3(a) and 3(b) show two representative tunneling spectra of the BLG: the relative intensities of the two spin-split $LL_{(0,1,+)}$ peaks are nearly the same at $v < 0$ (*i.e.*, the $LL_{(0,1,-)}$ is partially-filled), whereas they exhibit large asymmetry at $v = 0$ state (*i.e.*, the $LL_{(0,1,-)}$ is fully filled and the $LL_{(0,1,+)}$ is completely empty). In our experiment, the magnetic fields can induce redistribution of charges in graphene layers [25,33,46], therefore, the position of charge neutrality point of the BLG shifts slightly with increasing the magnetic fields. Figure 3(c) shows STS maps of the BLG as a function of magnetic fields. For the magnetic fields ranging from 5 T to 7 T, we obtain the filling factor $v < 0$; for the magnetic fields $B \geq 8$ T, the filling factor of the BLG becomes $v = 0$. The result in Fig. 3(c) clearly reveals the dependence of the relative intensity of the two spin-split peaks in the $LL_{(0,1,+)}$ on the filling states of the BLG. Similar phenomenon is also observed in another BLG sample (see Fig. S3 in Supplemental Material [28]). The observed large asymmetry of the intensities of the two spin-split $LL_{(0,1,+)}$ peaks is attributed to experimental evidence for the emergence of an orbital-polarized intermediate phase at the $v = 0$ filling state in the BLG [4,16,18]. There are three possible quantum phases at the $v = 0$ in the BLG depending on interlayer electric fields [6,14,16], as schematically shown in Figs. 3(d) and 3(e). At small interlayer electric fields, the BLG at the $v = 0$ will in a canted antiferromagnetic (CAF) state. At large interlayer electric fields, the BLG at the $v = 0$ is a layer (valley)-polarized state. Because different dependences of the $n = 0$ and 1 orbits on the electric fields, the single-particle picture predicts an intermediate $v = 0$ state between the CAF and layer-polarized states [4]. In this intermediate $v = 0$ state, there is a strong asymmetry of the two spin-split states at the $A_T$ sublattice, as shown in Fig. 3(e), which agrees well with that observed in our experiment (a large asymmetry of the two spin-split states at the $A_T$ sublattice is observed), as shown in Figs. 3(b), 3(c) and Fig. S4 [28]. In our BLG, the interlayer electric field is $E \sim 28$ mV, then a displacement field derived from $D = E/d$ is estimated as $D \sim 0.08$ V/nm ($d = 0.34$ nm is interlayer distance). In previous studies [4,16], the intermediate $v = 0$ state emerges with the displacement field ranging from 0.05 V/nm to 0.13 V/nm in high magnetic fields. Therefore, it is quite reasonable to observe the intermediate $v = 0$ state in our experiment.

The intermediate $v = 0$ state of the BLG is also directly imaged at atomic scale in our experiment by operating energy-fixed STS mappings, as shown in Fig. 4. Figures 4(b) and 4(c) show the atomic-scale STS maps measured at the energies of two spin-split $LL_{(0,1,+)}$ peaks for the intermediate $v = 0$ state. The STS map recorded at the left spin-split peak shows a clear triangular lattice with the dominant DOS on the $A_T$ site [Fig. 4(b)]. The STS map measured at the right spin-split peak exhibit a $A_T$-$B_B$ DOS pattern in the BLG atomic network [Fig. 4(c)]. To clearly compare the DOS distributions of the two spin-split peaks at the $v = 0$, we plot line cuts of the conductance maps of the two peaks in Fig. 4(d). Our measurement indicates that the DOS of the left spin-split peak is mainly localized on the $A_T$ atoms, and the DOS of the right spin-split peak is predominantly on the $A_T$ site and partly on the $B_B$ site. Such a result is well consistent with the scenario of the orbital-polarized intermediate $v = 0$ phase in the BLG. For comparison, we also measure the DOS distributions of the two spin-split peaks at the $v < 0$, as shown in Figs. 4(e)-4(h). Then, we observe a clear triangular lattice both at the spin-up and spin-down $LL_{(0,1,+)}$ peaks with the dominant DOS on the $A_T$ site and a lower DOS on the $B_T/A_B$ site. Such a result is quite reasonable because that the wavefunctions of electrons in the $LL_{(0,1,+)}$ are mainly localized on the $A_T$ atoms.

In summary, we study the broken symmetry LLs and the orbital-polarized intermediate $v = 0$ state in the BLG through high-resolution spectroscopic measurements. Rich LL splittings are observed in the high-magnetic-field tunneling spectra. We find that the spin and orbital splittings of the lowest LL depend on the filling state and shows an obvious increase at partial filling, indicating strong many-body interactions in the lowest LL of the BLG. At the $v = 0$ filling state, we obtain the tunneling spectroscopy of the intermediate phase and directly visualize this state at the atomic scale. Our experiment demonstrates that there are plenty of rooms to explore the exotic quantum phases in graphene systems by using STM and STS measurements.

**Acknowledgments**

This work was supported by the National Natural Science Foundation of China (Grant Nos. 11974050, 11804089, 11674029), the Natural Science Foundation of Hunan Province, China (Grant No. 2018JJ3025). L.H. also acknowledges support from the National Program for Support of Top-notch Young Professionals, support from "the Fundamental Research Funds for the Central Universities", and support from "Chang Jiang Scholars Program". L.J.Y. acknowledges support from the Fundamental Research Funds for the Central Universities.

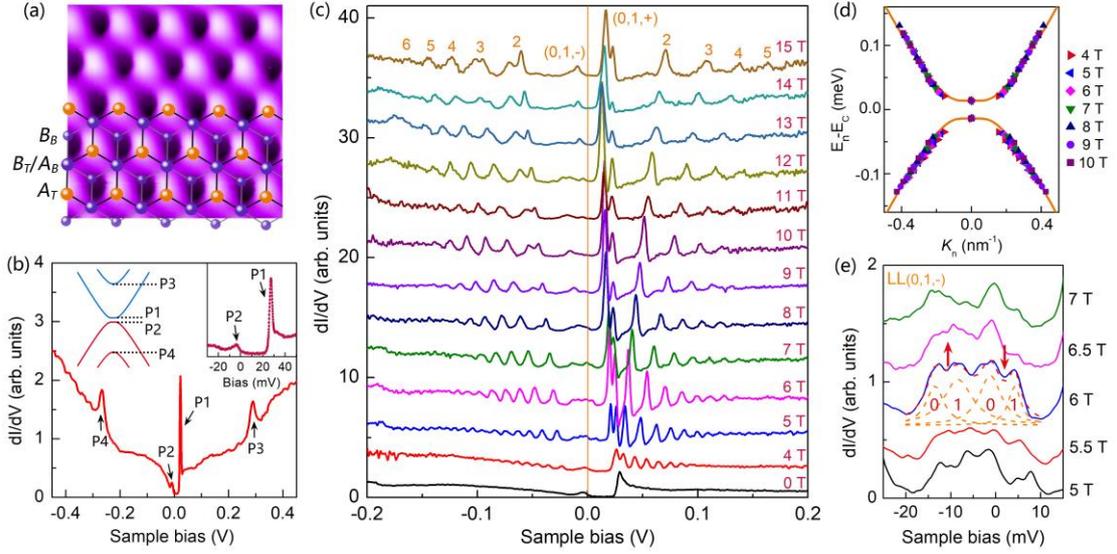

FIG. 1. (a) Atomic-resolution STM image (1.1 ×1.1 nm$^2$, $V_b$ = 77 mV, $I$ = 50 pA) of a decoupled *AB*-stacked BLG on graphite. The *AB* stacking configuration is superimposed with part of the image. Balls are carbon atoms, and black (grey) lines indicate C-C bonds in the top (bottom) layer. (b) Typical *dI/dV* spectrum for gapped BLG. Left inset shows the band structures of the gapped BLG. Right inset is a close-up of the band gap in the spectrum. (c) LL spectra of the BLG obtained from 0 T to 15 T. Curves are shifted vertically for clarity. The orange line represents the Fermi energy. (d) Effective band dispersion. The spots are data of $E_n$ (4-10 T) extracted from (c), which subtract the energy of charge neutrality point ($E_C$). The curves are tight-binding fits to the data. (e) *dI/dV* spectra around the Fermi level of the $n$ = (0,1,-) LL. Lorentz peak fitting (dashed curves) of the LL splitting is performed in 6 T as an example. Arrows (↑↓) denote spin-up and spin-down, 0 and 1 represent orbits.

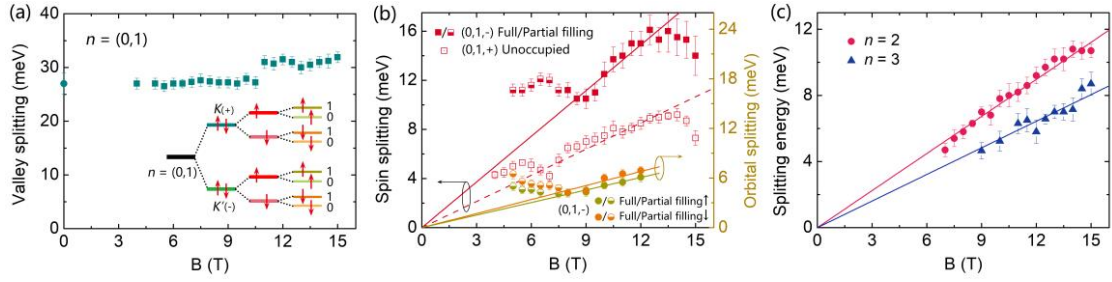

FIG. 2. (a) Valley splitting of the lowest $n = (0,1)$ LL as a function of magnetic field. The inset shows the sequence of symmetry-breaking for the $n = (0,1)$ LL. The 0 T data, which extracted from the band gap, is presented for comparison. (b) Spin splitting of $n = (0,1,-)$ and $(0,1,+)$ LLs and orbital splitting of $n = (0,1,-)$ LL as a function of magnetic field. Colored lines are linear fits to the data. (c) Splitting energy of $n = 2$ and 3 LLs in the hole-side.

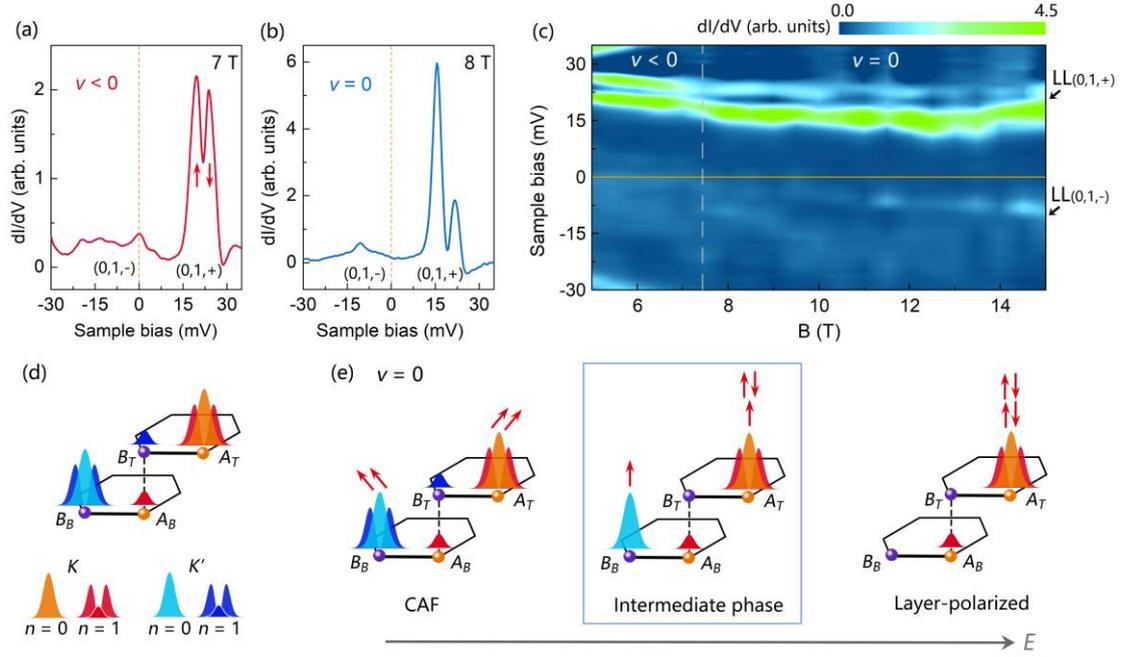

FIG. 3. *dI/dV* spectra of the lowest LL for *v* < 0 (a) and *v* = 0 (b) filling state taken at 7 T and 8 T, respectively. (c) *dI/dV* colour map of the lowest LL measured from 5 T to 15 T. The dashed line indicates the boundary between *v* < 0 and *v* = 0 states. (d) Schematic of the wavefunction distributions on the four atomic sites of BLG unit cell for the *n* = 0 and 1 LLs. *K* and *K'* denote two valleys of BLG. (e) Schematic of the CAF, intermediate phase and layer-polarized phase at the *v* = 0 filling state.

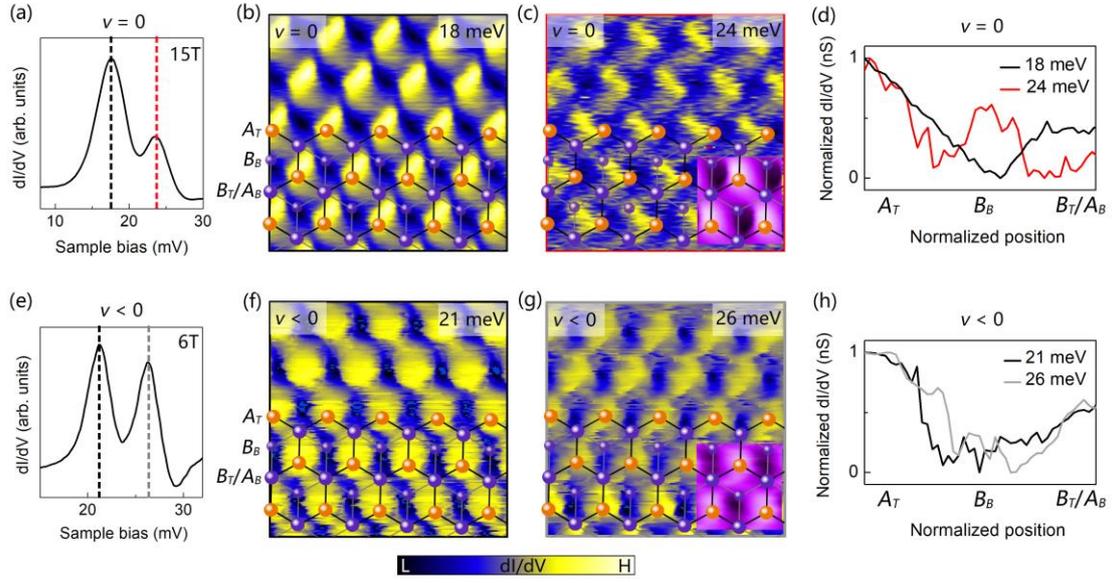

FIG. 4. (a) *dI/dV* spectrum of the split LL$_{(0,1,+)}$ taken at 15 T for *v* = 0. (b) and (c) STS spatial maps (1.1×1.1 nm$^2$) recorded at the peak energies of LL$_{(0,1,+)}$ as marked by black (18 meV) and red (24 meV) dashed lines in (a), respectively. (d) *dI/dV* spatial line cuts in the second-nearest $A_T$-$B_T$/$A_B$ direction taken from (b) and (c). (e) *dI/dV* spectrum of the split LL$_{(0,1,+)}$ taken at 6 T for *v* < 0. (f) and (g) STS spatial maps (1.1×1.1 nm$^2$) measured at energy positions indicated by black (21 meV) and grey (26 meV) dashed lines in (e), respectively. (h) STS line cuts in the next-nearest $A_T$-$B_T$/$A_B$ direction taken from (f) and (g). The insets in (c) and (g) show the simultaneously obtained topographic image superimposed with the atomic structure of BLG.